\input{aipcheck}

\documentclass[
    ,final            
  ]
  {aipproc}

\layoutstyle{6x9}

\begin{document}

\title{Game Theory Formulated on Hilbert Space}


\author{Taksu Cheon}
{
  address={Laboratory of Physics,
  		Kochi University of Technology,
  		Tosa Yamada, Kochi 782-8502, Japan},
  email={taksu.cheon@kochi-tech.ac.jp},
  thanks={This work was commissioned by the AIP}
}

\copyrightyear  {2006}

\begin{abstract}
We present a consistent formulation of quantum game theory 
that accommodates all possible strategies in Hilbert space.
The physical content of the quantum strategy is revealed as
a family of classical games representing altruistic game play supplemented by quantum interferences.  
Crucial role of the entanglement in quantum strategy is illustrated by an example of
quantum game representing the Bell's experiment.
\end{abstract}

\date{\today}

\maketitle

\section{Introduction}
The quantum game theory \cite{ME99, EW99, IQ05, FL05, IC06} has two aspects.
From one side, it is an extension of conventional game theory with
Hilbert space vectors and operators.
From the other side, it is an attempt to reformulate the description of
quantum information processing with the concept of payoff maximization.

In conventional game theory, strategies of players are represented by real-valued vectors,
and payoffs by real-valued matrices with no further specifications.  In quantum game theory,
they are replaced by complex {\it unitary} vectors and {\it Hemitian} matrices.  
It appears that the criterion of mathematical  beauty alone favors the latter over the former.
Since the space of classical strategies forms a subset of the entire quantum strategy space,
it is quite natural to regard the game theory formulated on Hilbert space as a logical extension of  
classical game theory.
It is tempting to imagine that, in search of natural extension, the quantum game theory
could have eventually been found irrespective to the discovery of quantum mechanics itself.
A crucial questions then arise: What is the {\it physical content} of quantum strategies?
Which part of  a quantum strategy is classically interpretable and which part purely quantum?  
Answers to these questions should  also supply a key to understand the mystery 
surrounding the ``quantum resolution''  of games with classical dilemmas \cite{BH01,EP02}.
Obviously, the answers to these questions are to be obtained only through a consistent
formulation of game strategies on Hilbert space.
When that is achieved, it can be used as a springboard to deal with
the second aspect of the quantum game theory; namely,
quantum games played with microscopic objects in states with full quantum superposition
and entanglement.  
In quantum information theory, concept of efficiency occasionally arises.  That would
supply the payoff function once we are able to
identify ``game players'' in the information processing.  It should then become possible
to reformulate the problem with the language of quantum games.

In this note, we formulate quantum strategies for classical games with 
{\it diagonal payoff matrices}, and clarify the classical and quantum contents
of the resulting payoff function.
We will discover two striking features in the results; the existence of a third party, and
the mixture of altruistic strategies.
We also sketch the game theoretic formulation of quantum information processing 
through an example of Bell's experiment.  We naturally recover the Tsirelson's limit.
\section{Game Strategy and Payoff on Hilbert Space}

We start by considering $n$-dimensional
Hilbert spaces ${\cal H}_A$ and ${\cal H}_B$ in which the strategies 
of the two players $A$ and $B$ are represented by vectors 
$ \left | \alpha \right>_A \in {\cal H}_A$ and $ \left | \beta \right>_B \in {\cal H}_B$. 
The space of {\it joint strategies} of the game is given by the 
direct product ${\cal H} ={\cal H}_A \times {\cal H}_B$.  A vector in ${\cal H}$ 
representing a joint strategy of the two players can be written \cite{CT06} as
\begin{eqnarray}
\label{jointst}
\left | \alpha, \beta; \gamma \right> 
= J(\gamma) \left | \alpha \right>_A \left | \beta \right>_B,
\end{eqnarray}
where the unitary operator $J(\gamma)$ provides quantum correlation 
({\it e.g.,} entanglement) for the separable 
states $ \left | \alpha \right>_A \left | \beta \right>_B$.
The two-body operator $J(\gamma)$ is  independent of the players' choice and is 
determined by a third party, which can be regarded as a {\it coordinator} of the game. 

Once the joint strategy is specified with $J(\gamma)$, the players are to receive the payoffs, 
which are
given by the expectation values of Hermitian operators $A$ and $B$:
\begin{eqnarray}
\label{QNash}
\Pi_A(\alpha, \beta; \gamma) 
&=& \left < \alpha, \beta; \gamma  | A  |  \alpha, \beta; \gamma  \right > ,
\\ \nonumber
\Pi_B(\alpha, \beta; \gamma) 
&=& \left < \alpha, \beta; \gamma  | B  |  \alpha, \beta; \gamma  \right > .
\end{eqnarray}
Both players try to optimize their strategy to gain the maximal payoff, 
and the result is 
the quantum version of the Nash equilibrium, where  
we have  
$(\alpha, \beta)= (\alpha^\star, \beta^\star)$ in the strategy space, 
at which point the payoffs separately attain the maxima as
\begin{eqnarray}
\label{AlBeNash}
\left. \delta_\alpha \Pi_A (\alpha, \beta^\star; \gamma)\right|_{ \alpha^\star}  = 0,
\quad
\left. \delta_\beta \Pi_B (\alpha^\star, \beta; \gamma) \right|_{\beta^\star} = 0,
\end{eqnarray}
under arbitrary variations in $\alpha$ and $\beta$.
We express the individual strategies in terms of orthonormal basis strategies
$\{ \left| i \right>\}$, $i = 1, ..., n$ which we regard as common to $A$ and $B$.
\begin{eqnarray}
\label{basis}
\left | \alpha \right>_A =  \sum_i \alpha_i \left | i \right>_A ,
\quad
\left | \beta \right>_B =     \sum_i \beta_i \left | i \right>_B ,
\end{eqnarray}
with complex numbers $\alpha_i$, $\beta_i$ 
normalized as $\sum_i \vert \alpha_i \vert^2 =  \sum_i \vert \beta_i \vert^2 = 1$.
We introduce the swap operator $S$ by
\begin{eqnarray}
S \left | i , j \right> =  \left | j , i \right>
\end{eqnarray}
for the states $\left | i ,  j \right> = \left | i \right>_A \left | j \right>_B$,
and then $S \left | \alpha, \beta \right> = \left | \beta, \alpha \right>$ 
for general separable states 
$\left | \alpha, \beta \right> =  \left | \alpha \right>_A \left | \beta \right>_B$ results.   
We further introduce operators $C$ and $T$ by
\begin{eqnarray}
C \left | i , j \right>  =  \left | {\bar i}, {\bar j} \right> ,
\quad
T \left | i , j \right>  =  \left | {\bar j}, {\bar i} \right> ,
\end{eqnarray}
where the bar represents the complimentary choice;
${\bar i} $ $=  (n-1)-i$.
The operator $C$ is the simultaneous renaming (conversion) of strategy for two players, 
and $T$ is the combination
$T = CS$.
These operators $\{S, C, T\}$ commute among themselves and 
satisfy 
\begin{eqnarray}
S^2=C^2=T^2 = I,
\nonumber \\ 
T=SC, S=CT, C=TS, 
\end{eqnarray}
where $I$ is the identity operator. 
They form the dihedral group $D_2$.

By defining  the {\it correlated payoff operators}
%
\begin{eqnarray}
{\cal A}(\gamma) 
= J^\dagger(\gamma)A J(\gamma) , \quad  
{\cal B}(\gamma) 
= J^\dagger(\gamma)B J(\gamma) ,
\end{eqnarray}
we have
$\Pi_A (\alpha, \beta; \gamma)$
$=\left< \alpha,\beta \right| {{\cal A}(\gamma)} \left| \alpha,\beta \right> $.  
We consider diagonal payoff matrices whose elements are given by 
%
\begin{eqnarray}
\label{diagAB}
\left< i', j' \right| A \left| i, j \right>
\!\!\! &=& \!\!\!    A_{ij} \delta_{i' i}\delta_{j' j},
\\ \nonumber
\left< i', j' \right| B \left| i, j \right>
\!\!\! &=& \!\!\!    B_{ij} \delta_{i' i}\delta_{j' j}.  
\end{eqnarray}
Observe that we have
\begin{eqnarray}
\Pi_A (\alpha, \beta; 0) = \sum_{i,j} x_i  A_{ij} y_j ,
\\ \nonumber
\Pi_B (\alpha, \beta; 0)= \sum_{i,j} x_i  B_{ij} y_j ,
\end{eqnarray} 
%
where $x_i = \vert \alpha_i\vert^2$ and $y_j = \vert \beta_j\vert^2$ are
the probability of choosing the strategies 
$ \left | i \right>_A$ and $ \left | j \right>_B$ respectively.  This means that, at $\gamma = 0$, 
our quantum game reduces to the classical game with the payoff matrix $A_{ij}$ 
under mixed strategies.  
\section{Altruistic Contents and Quantum Interferences in Quantum Games}

Let us now restrict ourselves to two strategy games $n = 2$.  
The unitary operator $J(\gamma)$ then admits the form,
\begin{eqnarray}
\label{Hparam1}
J(\gamma) =  \, e^{i \gamma_1 S / 2} e^{i \gamma_2 T / 2} ,
\end{eqnarray}
where $\gamma = (\gamma_1,  \gamma_2)$ are real parameters. 
Note that, on account of the relation $S+T-C=I$
valid for $n = 2$, only two operators are independent in the set $\{S, C, T\}$.
The correlated payoff operator $A(\gamma)$ is split into two terms 
\begin{eqnarray}
\label{Agam}
{\cal A}(\gamma) 
= {\cal A}^{\rm pc}(\gamma) + {\cal A}^{\rm in}(\gamma)
\end{eqnarray}
where ${\cal A}^{\rm pc}$ is the  ``pseudo classical'' term and 
${\cal A}^{\rm in}$ is the ``interference'' term given, respectively, by
\begin{eqnarray}
\label{Agam01}
{\cal A}^{\rm pc}(\gamma) 
\!\!\!&=&\!\!\! 
   \cos^2{ {\gamma_1\over 2}} A   +
 ( \cos^2{ {\gamma_2\over 2}} - \cos^2{ {\gamma_1\over 2}} ) S A S
  + \sin^2{ {\gamma_2\over 2}} C A C ,
\nonumber \\
%
{\cal A}^{\rm in}(\gamma) 
\!\!\!&=&\!\!\! 
    { {i } \over {2} } \sin \gamma_1(AS - SA)
 + { {i} \over {2} }  \sin \gamma_2(AT - TA) .
\end{eqnarray}
Correspondingly, the full payoff is also split into two contributions from
${\cal A}^{\rm pc}$ and ${\cal A}^{\rm in}$ as $\Pi_A=$
$\Pi_A^{\rm pc}+ \Pi_A^{\rm in}$.   To evaluate the payoff, we may choose both
$\alpha_0$ and $\beta_0$ to be real without loss of  generality, 
and adopt the notaions $(\alpha_0, \alpha_1)=(a_0, a_1 e^{i\xi})$ 
and $(\beta_0, \beta_1)=(b_0, b_1 e^{i\chi})$.  The outcome is
\begin{eqnarray}
\label{payof0}
& &
\Pi_A^{\rm pc}(\alpha,\beta; \gamma)
=
\sum_{i,j} { a_i^2 b_j^2  {\cal A}^{\rm pc}_{i j} }(\gamma) ,
\\ \nonumber
& &
\Pi_A^{\rm in}(\alpha,\beta; \gamma)
= - a_0 a_1 b_0 b_1 
[ G_+(\gamma) \sin(\xi + \chi) + G_-(\gamma)  \sin(\xi - \chi) ] ,
\quad
\end{eqnarray}
with 
${\cal A}^{\rm pc}_{i j} (\gamma) = \left< i, j \right| {\cal A}^{\rm pc}(\gamma)  \left| i, j \right>$ 
and 
\begin{eqnarray}
\label {FGH}
G_+(\gamma)  \!\!\!& = &\!\!\!  (A_{00}  -  A_{11}) \sin\gamma_2,
\\ \nonumber 
G_-(\gamma)   \!\!\!& = &\!\!\!  (A_{01}  -  A_{10})\sin\gamma_1.
\end{eqnarray}

A completely parallel expressions are obtained for the payoff matrix $B(\gamma)$
and the payoff $\Pi_B(\alpha,\beta; \gamma)$ for the player $B$.

Above split of the payoff shows that  the quantum game consists of two ingredients.  
The first is the pseudo classical ingredient associated with ${\cal A}^{\rm pc}(\gamma)$, 
whose form indicates that we are, in effect, 
simultaneously playing three different classical games, {\it i.e.,} the original classical
game $A$, and two types of ``converted'' games,
specified by diagonal matrices $SAS$ and $CAC$
with the mixture specified by given $\gamma_1$ and $\gamma_2$.   
Regarding $\gamma$ as tunable parameters, we see that the quantum game 
contains a {\it family} of classical games that includes the original game.  
The second ingredient of the quantum game is the purely quantum 
component  ${\cal A}^{\rm in}(\gamma)$, 
which occurs only when both of the two players adopt quantum strategies 
with $a_0 a_1 b_0 b_1 \ne 0$ and non-vanishing phases $\xi$ and $\chi$.   
The structure of $\Pi_A^{\rm in}$ suggests that this interference term cannot be simulated by 
a classical game and hence represents the {\it bona fide} quantum aspect.
We further look into the pseudo classical family to uncover its physical content.
To that end, we assume that one of the coordinator's parameters, $\gamma_2$ 
is zero.  We have
\begin{eqnarray}
\label{ABgamS}
{\cal A}^{\rm pc}(\gamma_1) 
\!\!&=&\!\!
   \cos^2 \frac{\gamma_1}{2} A   +  \sin^2  \frac{\gamma_1}{2} S A S
\\ \nonumber
{\cal B}^{\rm pc}(\gamma_1) 
\!\!&=&\!\!
   \cos^2 \frac{\gamma_1}{2} B   +  \sin^2 \frac{\gamma_1}{2} S B S
\end{eqnarray}
The meaning of these payoff matrices becomes evident by considering
a {\it symmetric game}, which is defined by requiring 
that  the payoffs are symmetric for two players, namely
$\Pi_A (\alpha, \beta; \gamma) = \Pi_B ( \beta, \alpha; \gamma)$.  
The game appears identical to both players $A$ and $B$.
In this sense, a symmetric game is {\it fair} to both parties.
It is easy to see that the condition of symmetry translates into the requirement $B = SAS$.
We then have, for a symmetric game,
\begin{eqnarray}
\label{ABgamSS}
{\cal A}^{\rm pc}(\gamma_1) 
\!\!&=&\!\!
   \cos^2 \frac{\gamma_1}{2} A   +  \sin^2  \frac{\gamma_1}{2} B
\\ \nonumber
{\cal B}^{\rm pc}(\gamma_1) 
\!\!&=&\!\!
   \cos^2 \frac{\gamma_1}{2} B   +  \sin^2 \frac{\gamma_1}{2} A .
\end{eqnarray}
This means that the pseudo classical game specified by modified rule 
${\cal A}(\gamma_1)$ and
${\cal B}(\gamma_1)$ can be interpreted as a game played with 
the mixture of {\it altruism}, or players' taking into account of other party's interest
along with their own self-interest \cite{CH03, CH05}. The degree of mixture of
altruism is controlled by 
the correlation parameter $\gamma_1$. 
It is a well known fact that altruistic behavior is widespread among primates that
lead social life.  It is also well known that the introduction of altruism ``solves'' such
long-standing problems as prisoner's dilemma, to which attempts for solution within
conventional game theories based solely on narrow egoistic self-interest has been
notoriously difficult \cite{AX84,TA87}. 
If we fix the first correlation parameter to be $\gamma_1=\pi/2$,  and assume
{\it T-symmetric} game $B = TAT$, we arrive at a parallel relation to (\ref{ABgamSS}),
thereby proving the fact that pseudo classical family is essentially made up of classical
games with altruistic modification specified by the coordinator's parameter $\gamma$. 

For detailed solutions of Nash equilibria with exhaustive classification 
according to the relative value of the payoff parameters, readers are referred 
to \cite{CT06, IT06, IC06}.

\section{Bell Experiment as a Quantum Game}

What we have done up to now amounts to ``quantizing'' 
classical games.  With the advent of nanotechnology, however, it is now possible to
actually set up a {\it game with quantum particles} as a laboratory experiment
that has no classical analogue.
For such quantum games, we have to allow arbitrary Hermitian payoff operators
$A$ and $B$, removing the restriction to diagonal ones, (\ref{diagAB}).
Without the diagonal condition, however, it turns out that the parametrization
of Hilbert  space ${\cal H}_A\times{\cal H}_B$ with the correlation operator
 (\ref{Hparam1}) is not 
completely valid. (It leaves certain relative phases between 
basis states fixed, which, for the case of diagonal payoff operators, does no harm.)
Instead, we resort to the scheme devised by Cheon, Ichikawa and Tsutsui \cite{CIT06} that
utilizes Schmidt decomposition
\begin{eqnarray}
\label{schmdR}
\left| \Psi(\alpha,\beta; \eta) \right> 
= U(\alpha) \otimes U(\beta) \left| \Phi(\eta) \right> ,
\end{eqnarray}
with ``initial'' correlated state
\begin{eqnarray}
\label{schmd}
 \left| \Phi(\eta) \right> 
= \cos{\frac{\eta_1}{2}} \left|0 0 \right>
 + e^{i\eta_2} \sin{\frac{\eta_1}{2}}  \left|1 1 \right> ,
\end{eqnarray}
and individual $SU(2)$ rotations $U(\alpha)$ and $U(\beta)$ that are controlled respectively by
player $A$ and $B$.  For definiteness we write
\begin{eqnarray}
U(\alpha) = 
 \pmatrix{
  \cos\frac{\theta_\alpha}{2}  & - e^{-i\varphi_\alpha}\sin\frac{\theta_\alpha}{2} \cr
   e^{i\varphi_\alpha}\sin\frac{\theta_\alpha}{2} &  \cos\frac{\theta_\alpha}{2} 
} .
\end{eqnarray}
The Schmidt state $\left| \Psi(\alpha,\beta; \eta) \right>$  covers 
{\it entire} Hilbert space ${\cal H}_A\times{\cal H}_B$.
Note that the coordinator's parameters $\eta_1$ and $\eta_2$ have definite
meaning as the measure of size and phase of {\it two-particle entanglement}.  

As an example of such quantum game, let us consider payoff operators
\begin{eqnarray}
A = B = \sqrt{2} \left(  \sigma_x\otimes\sigma_x + \sigma_z\otimes\sigma_z \right) .
\end{eqnarray}
This is nothing other than the measurement operator for Bell's experiment,
in which the projection of two spin $1/2$ particles specified by the state (\ref{schmdR}) 
are measured separately. Here we identify $\left| 0 \right>$ and  $\left| 1 \right>$ as
``up'' and ``down'' states of spin 1/2 along $z$ axis, namely
\begin{eqnarray}
\sigma_z \left| 0 \right> = \left| 0 \right>,
\quad 
\sigma_z \left| 1 \right> = - \left| 1 \right> .
\end{eqnarray}
The spin projection of the first particle is measured either
along positive $x$ axis (whose value, we call  $P_{1}$) 
or along positive $z$ axis (whose value is $P_{2}$) with random alternation.
The spin projection of the second particle is measured either along  the line $45$ degrees 
between positive $x$ and $z$ axes ($Q_{1}$), 
or along the line $45$ degrees between negative $x$  
and positive $z$ axes ($Q_{2}$), again in random alternation.  Suppose
that both players are interested in maximizing the quantity
\begin{eqnarray}
\Pi \equiv P_1Q_1-P_2Q_2.
\end{eqnarray}
We can easily show that  $\Pi$ is given by the common payoff
to $A$ and $B$ given by
\begin{eqnarray}
\Pi = 
\Pi_A(\alpha,\beta,\eta) 
= \Pi_B(\alpha,\beta,\eta)
= \left< \Psi(\alpha,\beta; \eta) \right| A \left| \Psi(\alpha,\beta; \eta) \right> .
\end{eqnarray}
The game now becomes a one of quantum coordination
between players $A$ and $B$ who both 
try to increase the common payoff $\Pi$
by respectively controlling the directions of spins with $U(\alpha)$ and $U(\beta)$.    
Considering the relation
\begin{eqnarray}
 \left< \Psi(\alpha,\beta; \eta) \right| A \left| \Psi(\alpha,\beta; \eta) \right> 
= \left< \Phi(\eta) \right| 
(U^\dagger(\alpha)\otimes U^\dagger(\beta) \,A\, U(\alpha)\otimes U(\beta)) 
\left| \Phi(\eta) \right> .
\end{eqnarray}
we can also restate the game as two players, receiving the correlated
two particle state $ \left| \Phi(\eta) \right>$, trying to maximize 
the common payoff $\Pi$ by rotating the directions of spin projection measurement:
The player $A$ applies a common rotation $U(\alpha)$ to the directions $P_1$ and $P_2$,
and the player $B$ applies  another common $U(\beta)$ to $Q_1$ and $Q_2$.
For a fixed set of entanglement parameters $(\eta_1, \eta_2)$, 
a straightforward calculation yields the Nash equilibrium that is specified by
\begin{eqnarray}
\theta_\alpha^\star = \theta_\beta^\star = {\it arbitrary},
\quad
\varphi_\alpha^\star =\varphi_\beta^\star =0 ,
\end{eqnarray}
for which, the Nash payoff is given by
\begin{eqnarray}
\Pi^\star(\eta_1,\eta_2)= \sqrt{2} ( 1+\sin{\eta_1}\cos{\eta_2} ) .
\end{eqnarray}
For particles with no entanglement, $\eta_1 = 0$, we obtain $\Pi^\star = \sqrt{2}$, which
is the known maximum for two uncorrelated spins. 
For particles with maximum entanglement, $\eta_1 = \pi/2$ and phase $\eta_2 = 0$,
we obtain the payoff $\Pi^\star = \sqrt{8}$, which is exactly on the Tsirelson's bound \cite{TS80}.
This reformulation of Bell's experiment should give a hint for the way toward 
more general game-theoretic reformulation of quantum information processing.

\begin{theacknowledgments}
This work has been partially supported by 
the Grant-in-Aid for Scientific Research of  Ministry of Education, 
Culture, Sports, Science and Technology, Japan
 under the Grant number 18540384.
\end{theacknowledgments}


\end{document}